\documentclass[12pt,times]{article}
 \usepackage[pdftex]{color,graphicx}
\usepackage{enumerate}
\usepackage{amsmath}
\usepackage{amssymb}
\usepackage{alltt}
\usepackage{setspace}
\usepackage{subfigure}
\usepackage{sidecap}
\usepackage{sectsty}
\partfont{\large}
\sectionfont{\large}
\subsectionfont{\small}
\usepackage[left=2cm,top=2cm,right=3cm,head=1cm,foot=1cm]{geometry}
\numberwithin{equation}{section}
\let\oldsqrt\sqrt
\def\sqrt{\mathpalette\DHLhksqrt}
\def\DHLhksqrt#1#2{%
\setbox0=\hbox{$#1\oldsqrt{#2\,}$}\dimen0=\ht0
\advance\dimen0-0.2\ht0
\setbox2=\hbox{\vrule height\ht0 depth -\dimen0}%
{\box0\lower0.4pt\box2}}
\newcommand{\al}{\alpha}
\newcommand{\g}{\gamma}
\newcommand{\e}{\varepsilon}
\newcommand{\ta}{\theta}

\newcommand{\G}{\Gamma}

\newcommand{\ph}{\varphi}
\newcommand{\da}{\dagger}
\newcommand{\pa}{\partial}

\newcommand{\la}{\mathcal L}
\newcommand{\ld}{\lambda}

\newcommand{\D}{\mathcal D}
\newcommand{\M}{\mathcal M}

\newcommand{\F}{\mathcal F}
\newcommand{\A}{\mathcal A}
\newcommand{\B}{\mathcal B}

\newcommand{\x}{\times}

\newcommand{\ml}{\left(\begin{matrix}}
\newcommand{\mr}{\end{matrix}\right)}

\newcommand{\U}{\mathcal U}

\newcommand{\w}{\omega}

\newcommand{\bra}{\langle}
\newcommand{\ket}{\rangle}
\newcommand{\tr}{\text{tr}}
\newcommand{\op}{\mathcal O}
\newcommand{\del}{\delta}
\newcommand{\Del}{\Delta}
\newcommand{\lrarrow}{\leftrightarrow}
\newcommand{\zb}{\mathbb Z}
\newcommand{\ka}{\kappa}
\newcommand{\VPMNS}{V_{\text{PMNS}}}
\newcommand{\VCKM}{V_{\text{CKM}}}

\newcommand{\re}{\text{Re}}
\newcommand{\im}{\text{Im}}

\newcommand{\half}{\tfrac{1}{2}}
\newcommand{\third}{\tfrac{1}{3}}
\newcommand{\fourth}{\tfrac{1}{4}}
\newcommand{\sixth}{\tfrac{1}{6}}

\newcommand{\s}{\sigma}

\newcommand{\msol}{\Del m^2_{\text{sol}}}
\newcommand{\matm}{\Del m^2_{\text{atm}}}

\begin{document}
\title{\Large \textbf{An $A_4\x \zb_4$ model for neutrino mixing}}
\author{Yoni BenTov$^{1}$, Xiao-Gang He$^{3,4,5}$, A. Zee$^{1,2}$}
\date{}
\maketitle
\begin{flushleft}
\textit{$^1$\,Department of Physics, University of California, Santa Barbara CA 93106}\\
\textit{$^2$\,Kavli Institute for Theoretical Physics, University of California, Santa Barbara CA 93106}\\
\textit{$^3$\,INPAC, Department of Physics and Shanghai Key Laboratory for Particle Physics and Cosmology, Shanghai Jiao Tong University, Shanghai, China}\\
\textit{$^4$\,CTS, CASTS and Department of Physics, National Taiwan University, Taipei, Taiwan}\\
\textit{$^5$\,Department of Physics, National Tsing Hua University, and National Center for Theoretical Sciences, Hsinchu, 300, Taiwan}
\end{flushleft}
\begin{abstract}
The $A_4\x U(1)$ flavor model of He, Keum, and Volkas is extended to provide a minimal modification to tribimaximal mixing that accommodates a nonzero reactor angle $\ta_{13} \sim 0.1$. The sequestering problem is circumvented by forbidding superheavy scales and large coupling constants which would otherwise generate sizable RG flows. The model is compatible with (but does not require) a stable or metastable dark matter candidate in the form of a complex scalar field with unit charge under a discrete subgroup $\zb_4$ of the $U(1)$ flavor symmetry.
\end{abstract}
\section{Introduction}\label{sec:intro}
The replication of fermion generations along with small quark mixing angles and large neutrino mixing angles may suggest that the matter content of the Standard Model (SM) transforms nontrivially under a horizontal flavor symmetry group, $G_F$. If this flavor group $G_F$ were a continuous global symmetry, then spontaneously breaking it would induce massless Goldstone bosons in the low-energy spectrum, which are difficult to reconcile with experiment. One might therefore imagine that either $G_F$ is actually gauged at high energy \cite{gauged flavor} or that the global family symmetry is actually a discrete group \cite{discrete flavor} rather than a continuous Lie group.
\\\\
The flavor group 
\begin{equation}\label{eq:GF}
G_F = A_4 \x U(1)
\end{equation}
\\
was originally proposed for lepton masses by E. Ma and G. Rajasekaran \cite{A4ma} and subsequently studied by many authors. The $U(1)$ is essentially lepton number,\footnote{To avoid generating a Goldstone mode upon spontaneously breaking the $U(1)$, it is convenient to introduce a new SM-singlet scalar field charged under only a discrete subgroup $\zb_n$ with a self-interaction that explicitly breaks the $U(1)$. We address this issue in Section~\ref{sec:DM}.} but where heavy gauge-singlet antineutrinos are assigned $U(1)$ charge zero instead of $-1$.
\\\\
In this paper we will extend a particular model by He, Keum, and Volkas (HKV) \cite{A4hkv} which uses the same symmetry group Eq.~(\ref{eq:GF}). A supersymmetric model containing many of the same leading-order features was proposed earlier by Babu, Ma, and Valle \cite{A4bmv}. Their approach to generate a realistic neutrino mixing matrix from the leading order result is to break $G_F$ at high energy and have the renormalization group generate a realistic neutrino mixing matrix at low energy; here we will take the opposite approach of attempting to break $G_F$ at as low a scale as possible, so that renormalization group corrections are negligible. We also do not require any use of supersymmetry.
\\\\
The model proposes that, to leading order, the CKM matrix is the identity, and the charged lepton and neutrino mass matrices are ``form diagonalized" by unitary matrices that result in a tribimaximal neutrino mixing matrix,
\begin{equation}\label{eq:VTB}
\VPMNS = V_{\text{TB}} \equiv \ml -\frac{2}{\sqrt6}&\frac{1}{\sqrt3}&0\\ \frac{1}{\sqrt6}&\frac{1}{\sqrt3}&\frac{1}{\sqrt2}\\ \frac{1}{\sqrt6}&\frac{1}{\sqrt3}&-\frac{1}{\sqrt2} \mr\;,
\end{equation}
independently of the values of the charged lepton and neutrino masses. The main theoretical challenge for this model is that the charged fermion mass matrices are obtained using an $A_4$ triplet vacuum expectation value (VEV) pointing in the direction $(1,1,1)$, while the neutrino mass matrix is obtained using an $A_4$ triplet VEV pointing in the direction $(0,1,0)$. The former leaves a $\zb_3$ subgroup of $A_4$ unbroken, while the latter leaves a $\zb_2$ subgroup of $A_4$ unbroken. The two groups do not commute, and so a generic scalar potential will spoil the two vacuum alignments. The authors called this the ``sequestering problem" and proposed one solution based on low-energy supersymmetry.
\\\\
In this paper we address the sequestering problem by embedding the HKV model in a framework which has no superheavy scales, and therefore does not generate large renormalization group corrections to quadratic scalar couplings. Thus in this model the sequestering problem is solved by showing that it is not a problem to begin with: if we set various quartic couplings in the potential to be negligible at some high energy scale, they will remain small at low energy simply because the separation of scales involved is not large. 
\section{Leading order masses and mixing}\label{sec:charged fermions}
The leading order results for quark mixing and neutrino oscillations are the same as in the HKV model, so here we will only review the field transformation properties under $SU(3)\!\x\! SU(2)\!\x\! U(1)\!\x\! A_4\! \x\! U(1)_X$ and the resulting expressions for the mass matrices.
\\\\
The SM fermion fields $q, \bar u, \bar d, \ell, \bar e$ and the SM-singlet antineutrino fields $N$ transform as:
\begin{align}\label{eq:fermions}
&q \sim (3,2,+\sixth; 3,0)\;,\;\; \bar d \sim (\bar 3,1,+\third; 1+1'+1'',0)\;,\;\; \bar u \sim (\bar 3,1,-\tfrac{2}{3};1+1'+1'',0) \nonumber\\
&\ell \sim (1,2,-\half; 3,+1)\;,\;\; \bar e \sim (1,1,+1; 1+1'+1'',-1)\;,\;\; N \sim (1,1,0; 3, 0)\;.
\end{align}
There are four $SU(2)\x U(1)$ Higgs doublets, $\{\Phi_a\}_{a\,=\,1}^3$ (where $a = 1,2,3$ labels the triplet representation of $A_4$) and $\phi$:
\begin{align}\label{eq:higgs doublets}
&\Phi \sim (1,2,-\half; 3,0)\;,\;\; \phi \sim (1,2,-\half; 1, +1)\;.
\end{align}
If we assume the $A_4$ VEV alignment $\bra\Phi_1^0\ket = \bra\Phi_2^0\ket = \bra\Phi_3^0\ket \equiv \tfrac{1}{\sqrt2}v\,e^{\,i\ta_\Phi}$, the mass matrices for the down quarks, up quarks, and charged leptons are of a common form:
\begin{equation}
M_f = \tfrac{1}{\sqrt2}Y_f v\,e^{\,i\ta_\Phi}\;,\;\; Y_f = \ml y_f &y_f'&y_f''\\ y_f&\w\, y_f'&\w^* y_f'' \\ y_f&\w^*y_f'&\w\,y_f'' \mr
\end{equation}
Here the index $f = d,u,e$ labels the three sectors of charged fermions $(Q = +\third, -\tfrac{2}{3}, -1$, respectively) in the SM.
\\\\
Thus the mass terms
\begin{equation}\label{eq:fermion mass terms}
\la_f = -(f_1,f_2,f_3)M_f \ml \bar f_1\\ \bar f_{1''}\\ \bar f_{1'} \mr+h.c.
\end{equation}
are diagonalized by the unitary transformation $f = \U_L^f f_{\text{mass}}$, where
\begin{equation}\label{eq:U_L^f}
\U_L^f = \tfrac{1}{\sqrt3}\ml e^{\,ia_f}&e^{\,ib_f}&e^{\,ic_f}\\e^{\,ia_f}&\w^*e^{\,ib_f}&\w\,e^{\,ic_f}\\ e^{\,ia_f}&\w\,e^{\,ib_f}&\w^*e^{\,ic_f} \mr\;.
\end{equation}
The phase angles $a,b,c$ are chosen to make the physical fermion masses real and positive. In Eq.~(\ref{eq:fermion mass terms}) the subscripts $1,2,3$ label the $A_4$ triplet representation, while the subscripts $1,1'',1'$ label the different $A_4$ singlet representations [see Eq.~(\ref{eq:fermions})].
\\\\
In this basis the Yukawa matrix is
\begin{equation}
(\U_L^f)^TY_f = \text{diag}(\sqrt3 y_f\,e^{\,ia_f},\sqrt3 y_f'\,e^{\,ib_f},\sqrt3 y_f''\,e^{\,ic_f})
\end{equation}
so the mass of each fermion is an arbitrary coupling constant times a VEV $v$, just as in the SM.
\\\\
The quark mixing matrix is $V_q \equiv (\U_L^d)^\da \U_L^u$. Putting this in the standard form $V_q \equiv \mathcal K_q\VCKM \mathcal P_q$, we find $\VCKM = I$. Thus this model explains why quark mixing is small; the realistic CKM angles are presumed to arise from higher order corrections, or from an extension to the theory. In this work we focus on neutrino mixing and do not pursue quark mixing any further. Consequently, from now on we suppress the $SU(3)$ quantum number when expressing group transformation properties of the fields.
\\\\
As emphasized by HKV, the fermion masses are totally arbitrary while the left-handed mixing matrices are, up to unphysical phases, fixed purely by group theory, in contrast to the alternative possibility that the mixing angles might be related to ratios of masses.
\section{Neutrino masses}\label{sec:neutrinos}
The leading contribution to neutrino masses arises through the seesaw mechanism. An $A_4$ triplet of SM-singlet real scalars
\begin{equation}
S \sim (1,1,0; 3,0)
\end{equation}
with an $A_4$ vacuum alignment $\bra S_1\ket = \bra S_3\ket = 0$, $\bra S_2\ket \equiv v_S \neq 0$ (with $v_S$ having either sign) gives a Majorana mass matrix for $N$:
\begin{equation}\label{eq:maj mass for N}
M_N = m_N\ml 1&0&x\\0&1&0\\x&0&1 \mr\;,\;\; x \equiv \frac{y_Nv_S}{m_N}\;.
\end{equation}
Here $m_N$ is a bare Majorana mass term, and $y_N$ is the Yukawa coupling of $S$ to $(NN)_{3_S}$. By rephasing the $N$ fields we can take $m_N$ real and positive without loss of generality. In this case, the parameter $x$ is in general a complex number with an undetermined phase \cite{A4hkv}. The Dirac mass term arises from $\la_{\text{Yuk}}^\nu = y_\nu\phi^\da \ell N$ and is proportional to the identity:
\begin{equation}\label{eq:dirac mass}
M_D = m_D\,I\;,\;\; m_D = \tfrac{1}{\sqrt2}y_\nu v_\phi\,e^{-i\ta_\phi}
\end{equation}
where $\tfrac{1}{\sqrt2}v_\phi\,e^{\,i\ta_\phi} \equiv \bra\phi^0\ket$. 
In the seesaw limit the light neutrino Majorana mass matrix is
\begin{equation}\label{eq:seesaw mass matrix}
M_\nu \approx -M_DM_N^{-1}M_D^T = -\,\frac{m_D^2}{m_N}\,\frac{1}{1-x^2}\ml 1&0&-x\\0&1-x^2&0\\-x&0&1 \mr\;.
\end{equation}
This is diagonalized in the form $(\U_L^\nu)^TM_\nu\, \U_L^\nu = D_\nu \equiv \text{diag}(m_1, m_2, m_3)$ by the unitary matrix
\begin{equation}\label{eq:U_L^nu}
\U_L^\nu = \tfrac{1}{\sqrt2}\ml e^{\,i\al/2}&0&-e^{\,i\g/2}\\0&\sqrt2\,e^{\,i\beta/2}&0\\e^{\,i\al/2}&0&+e^{\,i\g/2} \mr
\end{equation}
where the phase angles $\al,\beta,\g$ are chosen such that the physical neutrino masses $m_i$ are real and positive:
\begin{equation}\label{eq:neutrino masses}
m_1 = \frac{m_\nu}{|1+x|}\;,\;\;m_2 = m_\nu\;,\;\; m_3 = \frac{m_\nu}{|1-x|}
\end{equation}
where $m_\nu = |m_D^2/m_N|$. Neutrino oscillations constrain the mass-squared splittings $\matm \approx m_3^2-m_2^2$ and $\msol \approx m_2^2-m_1^2$ but not the overall scale $m_\nu$. Hence to compare with experiment one computes the ratio \cite{updatedreview}
\begin{equation}\label{eq:R data}
\xi \equiv \sqrt{|R|} = 5.29 \;\text{ to }\; 6.22 \qquad\text{where}\qquad R \equiv \frac{\matm}{\msol}\;. 
\end{equation}
This model has been shown to be compatible with Eq.~(\ref{eq:R data}) for both the ``normal" mass ordering $m_1 < m_2 < m_3$ ($\matm > 0$) and the ``inverted" mass ordering $m_3 < m_1 < m_2$ ($\matm < 0$) \cite{geometric nu mix}.
\\\\
The neutrino mixing matrix is
\begin{equation}
V_\nu \equiv (\U_L^e)^\da \U_L^\nu = \ml \tfrac{2}{\sqrt6}\,e^{i(\frac{\al}{2}-a_e)}&\tfrac{1}{\sqrt3}\,e^{\,i(\frac{\beta}{2}-a_e)}&0\\ -\frac{1}{\sqrt6}\,e^{\,i(\frac{\al}{2}-b_e+\frac{2\pi}{3})}&\frac{1}{\sqrt3}\,e^{\,i(\frac{\beta}{2}-b_e+\frac{2\pi}{3})}&-\frac{1}{\sqrt2}\,e^{\,i(\frac{\g}{2}-b_e+\frac{\pi}{6})}\\ -\frac{1}{\sqrt6}\, e^{\,i(\frac{\al}{2}-c_e-\frac{2\pi}{3})} & \frac{1}{\sqrt3}\,e^{\,i(\frac{\beta}{2}-c_e-\frac{2\pi}{3})}&-\frac{1}{\sqrt2}\,e^{\,i(\frac{\g}{2}-c_e-\frac{\pi}{6})} \mr\;.
\end{equation}
Thus, putting this in the standard form $V_\nu \equiv \mathcal K_\nu \VPMNS \mathcal P_\nu$ reproduces the tribimaximal mixing matrix of Eq.~(\ref{eq:VTB}): $\VPMNS = V_{\text{TB}}$.
\\\\
The recent measurement of a nonzero $V_{e3} \sim 0.1$ shows that this $\VPMNS$ cannot be exact at low energy. Just as for the leading order prediction $\VCKM \approx I$, this model predicts that $\VPMNS \approx V_{\text{TB}}$ at leading order, subject to small but nontrivial corrections.
\\\\
In this work we will use the fact that the $\zb_3$ and $\zb_2$ subgroups of $A_4$ do not commute to generate a realistic neutrino mixing matrix. We propose a model in which the $\zb_2$ remains a good approximate symmetry of the neutrino sector, but corrections due to the breaking of $\zb_3$ feed into the neutrino mass matrix and generate the nonzero $V_{e3}$ required to fit data. An alternative possibility is to perturb the theory around a vacuum that does not respect the $\zb_3$ symmetry, and use the corrections to $\zb_2$ in order to generate a nonzero $V_{e3}$.
\section{Scalar potential}\label{sec:potential}
As explained previously, we would like the charged lepton masses to arise from a VEV $\bra\Phi^0\ket \propto (1,1,1)$, which leaves invariant a $\zb_3$ subgroup of $A_4$, and we would like the neutrino masses to arise from a VEV $\bra S\ket \propto (0,1,0)$, which leaves invariant a $\zb_2$ subgroup of $A_4$. These two subgroups do not commute, and so the most general potential does not admit this configuration. This is called the sequestering problem.
\\\\
In the absence of a more elaborate framework, such as extra dimensions or supersymmetry, it is unlikely that this problem can be solved completely, but it can be sufficiently mitigated in a technically natural way. The general philosophy is to declare by fiat that certain couplings are small at a particular energy scale, and then to construct a model for which these couplings do not flow substantially under the renormalization group.\footnote{This is not an unfamiliar concept. In the SM, we declare by fiat that the electron Yukawa coupling is a dimensionless number of order $10^{-6}$. The reason we are able to do so consistently is that a small value for this coupling is protected by chiral symmetry from receiving large corrections. In other words, we fix a small value for a dimensionless coupling and show that it does not flow significantly under RG.}
\\\\
To make sure no superheavy scales are required, we need to bring down the scale of $m_N$ from the usual seesaw scale; for example, if $m_N \sim$ TeV then the neutrino mass scale $m_\nu \sim 10^{-2}$ eV can be obtained if $m_D \sim m_e \sim 10^{-1}$ MeV \cite{TeVseesaw}. In contrast, the simplest seesaw model has $m_N \sim M_{\text{GUT}} \sim 10^{12}$ TeV and $m_D \sim v \sim 10^{-1}$ TeV. Since the Dirac mass $m_D$ arises from the Higgs field $\phi$ [see Eq.(\ref{eq:dirac mass})], we want to make the VEV of $\phi$ small while keeping its mass above the weak scale.
\\\\
The problem of having heavy Higgs doublets with small VEVs was solved independently by Ma \cite{maTeV} and by Porto and Zee \cite{PH} and generalized by Grimus, Lavoura, and Radov\v{c}i\'{c} \cite{typeIIseesawhiggs}. In the present case we introduce a $G_{\text{SM}}\x A_4$-invariant complex scalar field $Y$ which carries $-2$ units of $U(1)_X$ charge:
\begin{equation}\label{eq:Y}
Y \sim (1,0; 1,-2)
\end{equation}
and a second $A_4$-invariant Higgs doublet
\begin{equation}\label{eq:phi'}
\phi' \sim (2,+\half; 1,+1)
\end{equation}
with the same $U(1)_X$ charge as $\phi$ but with opposite hypercharge [see Eq.~(\ref{eq:higgs doublets})]. The dimension-2 operator $\phi_i\e^{ij}\phi_j' \sim (1,0; 1,+2)$ is invariant under $G_{\text{SM}}\x A_4$ and can couple to $Y$ to form an invariant dimension-3 operator. Consider the following scalar potential:
\begin{align}\label{eq:private higgs type potential}
V(Y,\phi,\phi') &= M_Y^2Y^\da Y+M_\phi^2\phi^\da\phi+M_{\phi'}^2\phi'^\da\phi'+\ld_Y(Y^\da Y)^2+\ld_{\phi'}(\phi'^\da\phi')^2 \nonumber\\
&-(\mu Y\phi_i\e^{ij}\phi_j'+h.c.)+...
\end{align}
Here the ellipses contain all renormalizable terms consistent with the symmetry $G_{\text{SM}}\x G_F$ whose roles are assumed subdominant in what follows. We take $M_Y^2 < 0$ and $M_{\phi'}^2 < 0$ as usual, but $M_\phi^2 > 0$. When $Y$ and $\phi'^0$ obtain VEVs, they induce a VEV for $\phi^0$ according to the relation
\begin{equation}\label{eq:PH mass relation}
v_\phi = \frac{\mu \bra Y\ket v_{\phi'}}{M_\phi^2}\,+...
\end{equation}
where $v_\phi \equiv \sqrt2\bra\phi^0\ket$ and $v_{\phi'} \equiv \sqrt2\bra\phi'^0\ket$ are assumed real and positive for simplicity. Since there is no large separation of scales in this model, it is a self-consistent (though arbitrary) assumption to declare that the contributions denoted by ``..." in Eq.~(\ref{eq:PH mass relation}), which arise from quartic interactions in the Lagrangian, are subleading. If we can drop those contributions, then $|\bra\phi^0\ket| \sim 1/M_\phi^2$. Thus the VEV is inversely proportional to the square of the mass. For example, if $\mu \sim v_{\phi'} \sim 10^{-1}$ TeV and $\bra Y\ket \sim$ TeV, then
\begin{equation}\label{eq:PH mass}
M_\phi \sim \left( \frac{\text{MeV}}{v_\phi}\right)^{1/2}\!\!\!\times 10^2\,\text{TeV}\;.
\end{equation}
In this way we will assume that the $SU(2)\x U(1)$ doublet $\phi$ is the heaviest particle in the theory and thereby specify the values of all coupling constants at the scale $\M = M_\phi \sim 10^2$~TeV. This is much lower than the superheavy scales that often appear, e.g. $M_{\text{GUT}} \sim 10^{12}$ TeV, and so the renormalization group running to scales far below $\M$ will not be significant, and the non-commuting $\zb_3$ and $\zb_2$ subgroups of $A_4$ can remain sequestered up to small corrections. 
\section{Minimal modification to tribimaximal mixing}\label{sec:tribi mod}
Consider the case for which $\zb_2 = \{I,r_2\}$ is unbroken but $\zb_3 = \{I,c,a\}$ is broken by small perturbations that feed into the neutrino mass matrix. Since $\nu_1$ and $\nu_3$ flip sign under $\zb_2$, the mass terms $\nu_1\nu_2$ and $\nu_2\nu_3$ are still approximately zero. However, now that $c: \nu_{1,2,3} \to \nu_{2,3,1} \to \nu_{3,1,2}$ is broken, the terms $\nu_1\nu_1$ and $\nu_3\nu_3$ are no longer forced to be equal. The neutrino mass matrix in the $G_F$-basis is now of the form \cite{zeeTBmod}
\begin{equation}\label{eq:general modified M_nu}
M_\nu = \ml \al-\e&0&\beta\\ 0&\g&0\\ \beta&0&\al+\e \mr\;.
\end{equation}
If we assume that, to a good approximation, the charged lepton mass matrix remains diagonalized by the unitary matrix studied previously\footnote{Note that this $\zb_2$ leaves the $A_4$ singlets $1,1',1''$ invariant. This can be seen from decomposing the product of two $A_4$ triplets into irreducible representations: if $u,v \sim 3$, then the three singlets are of the form $u_1v_1+\w^{p_1}u_2v_2+\w^{p_2}u_3v_3$ with $p_1,p_2 = 0,1,2$. Since $u_1v_1$, $u_2v_2$, and $u_3v_3$ do not transform under $\zb_2$, neither does any linear combination of them.}, then the neutrino mixing matrix is a one-parameter deviation from tribimaximal mixing with the middle column $\propto (1,1,1)$ unchanged, which implies that a nonzero reactor angle $\ta_{13}$ is generated. To leading order in $\e$, we have
\begin{equation}\label{eq:Ve3 general form}
|V_{e3}| \approx \frac{|\e|}{\sqrt6|\beta|}\;.
\end{equation}
To search for a suitable high-energy model that will result in the desired perturbation, we first study the low-energy effective interactions which are invariant under $G_{\text{SM}} \x A_4$ between $\Phi \sim (2,-\half; 3,0)$ and the lepton doublets $\ell \sim (2,-\half; 3,+1)$ . Consider the dimension-5 interactions
\begin{equation}\label{eq:dim-5 lagrangian}
\la_{\text{dim-5}} = c_1\op_1+c_{1'}\op_{1'}+c_{1''}\op_{1''}+c_3\op_3+h.c.
\end{equation}
where:
\begin{align}
&\op_1 = (\Phi_1^\da\Phi_1^\da+\Phi_2^\da\Phi_2^\da+\Phi_3^\da\Phi_3^\da)(\ell_1\ell_1+\ell_2\ell_2+\ell_3\ell_3)\;,\\
&\op_{1'} = (\Phi_1^\da\Phi_1^\da+\w^*\Phi_2^\da\Phi_2^\da+\w\,\Phi_3^\da\Phi_3^\da)(\ell_1\ell_1+\w\,\ell_2\ell_2+\w^*\ell_3\ell_3)\;,\\
&\op_{1''} = (\Phi_1^\da\Phi_1^\da+\w\,\Phi_2^\da\Phi_2^\da+\w^*\Phi_3^\da\Phi_3^\da)(\ell_1\ell_1+\w^*\ell_2\ell_2+\w\,\ell_3\ell_3)\;,\\
&\op_{3} = (\Phi_2^\da\Phi_3^\da,\Phi_3^\da\Phi_1^\da,\Phi_1^\da\Phi_2^\da)\cdot(\ell_2\ell_3,\ell_3\ell_1,\ell_1\ell_2)\;.
\end{align}
These operators are invariant under $G_{\text{SM}}\x A_4$ but break $U(1)_X$ by two units. If we couple any of these to a $G_{\text{SM}}\x A_4$-invariant complex scalar $Y$ with $U(1)_X$ charge $-2$, then these operators are promoted to dimension-6, and the coefficients $c_i$ are proportional to $\bra Y\ket$.
\\\\
Consider the case $c_3 = 0$ and $\bra\Phi_1^0\ket = \bra\Phi_3^0\ket \neq \bra\Phi_2^0\ket$. Defining for convenience the overall scale of the VEVs and the small splitting as
\\
\begin{equation}\label{eq:a and delta}
a \equiv 3\bra\Phi_1^{0\da}\ket^2\;\;,\;\;\;\;\del \equiv \bra\Phi_2^{0\da}\ket^2-\bra\Phi_1^{0\da}\ket^2
\end{equation}
we find:
\begin{align}
c_1\op_1+c_{1'}\op_{1'}+c_{1''}\op_{1''} &= [ac_1+(c_1+\w^*c_{1'}+\w\,c_{1''})\del]\, \ell_1\ell_1 \nonumber\\
&+[ac_1+(c_1+c_{1'}+c_{1''})\del]\, \ell_2\ell_2 \nonumber\\
&+[ac_1+(c_1+\w\,c_{1'}+\w^*c_{1''})\del]\, \ell_3\ell_3\;.
\end{align}
The perturbation by $\del$ recovers a neutrino mass matrix in the form of Eq.~(\ref{eq:general modified M_nu}) with:
\begin{equation}
\e = i\,\tfrac{\sqrt3}{2}(c_{1'}-c_{1''})\del\;.
\end{equation}
Thus we seek a model in which $c_1 \sim c_{1'} \sim c_{1''} \neq 0$ and $c_3 \approx 0$, and in which $\bra\Phi^0\ket \propto (1,1,1)$ is perturbed in the form $\del\bra\Phi^0\ket \propto(\x,0,\x)$. In order to have the flavor-dependent\footnote{We are not working in the $e,\mu,\tau$ basis; we use the word ``flavor" here to mean that the perturbation depends on the $A_4$ triplet label $1,2,3$. This usage is consistent with the modern understanding of ``flavor" as meaning ``copy of representation of $G_{\text{SM}}$" -- in other words, copies of identical covariant derivatives -- rather than having any a priori relation to the charged lepton mass basis.} perturbation by $\del$ not be overwhelmed by the flavor-independent overall shift by $a$ [Eq.~(\ref{eq:a and delta})], we could attempt to construct a model in which $c_1 \ll c_{1'} \sim c_{1''}$. 
\section{Model for nonzero $V_{e3}$}\label{sec:Ve3}
One simple way to realize the effective operators from the previous section is to introduce $SU(2)\x U(1)$ triplets, $\Del, \Del', \Del''$, whose transformation properties under $SU(2)\x U(1) \x A_4 \x U(1)_X$ are\footnote{We express the $SU(2)$ components of the complex triplet $\Del$ as a 2-by-2 symmetric matrix:
\begin{equation}
\Del_{ij} = \ml \Del^0&\tfrac{1}{\sqrt2}\Del^-\\ \tfrac{1}{\sqrt2}\Del^-&\Del^{--} \mr
\end{equation}
where the superscript denotes the electric charge.}:
\begin{equation}\label{eq:higgs triplets}
\Del \sim (3,-1; 1,+2)\;,\;\; \Del' \sim (3,-1; 1', +2)\;,\;\; \Del'' \sim (3,-1; 1'',+2)\;.
\end{equation}
The field $\Del^\da$ can couple to $(\ell\ell)_1$, the field $\Del'^\da \sim (1')^* = 1''$ to $(\ell\ell)_{1'}$, and the field $\Del''^\da$ to $(\ell\ell)_{1''}$. The product $(\Del^\da)^{ij} (\Phi_i\Phi_j)_1$ is invariant under $SU(2)\x U(1)\x A_4$, but it carries charge $+2$ under $U(1)_X$. Fortunately the field $Y$ in Eq.~(\ref{eq:Y}) has just the required property to allow the dimension-4 interactions $Y^\da \Del^\da (\Phi\Phi)_1$, $Y^\da \Del'^\da (\Phi\Phi)_{1'}$, and $Y^\da \Del'^\da (\Phi\Phi)_{1''}$.
\\\\
First consider the following Lagrangian for $\Del$:
\begin{equation}\label{eq:Delta lagrangian}
\la_\Del = -M_\Del^2\tr(\Del^\da\Del)+\left\{(\Del^{\da})^{ij}\left[\mathcal A(\ell_i\ell_j)_1+\mathcal B^* Y^\da(\Phi_i\Phi_j)_1 \right] +h.c. \right\}+...
\end{equation}
Here $\mathcal A$ and $\mathcal B$ are couplings, and the ellipses denote other terms which are not important for the present discussion. We imagine that\footnote{The notation ``$a\lesssim b$" is to be interpreted as ``$a$ can be somewhat less than or possibly comparable to $b$". Also, one could instead take $M_\phi \lesssim M_\Del$, which would amount to pushing the scale at which $G_F$ is imposed up to $\M = M_\Del$. As long as this is still substantially smaller than the usual $M_{\text{GUT}}$, the main results of this paper remain unchanged. See Appendix~\ref{sec:RG} for a toy model.} $v < M_\Del \lesssim M_{\phi} \sim 10^2$ TeV. At scales far below $M_\Del$, the path integral for $\Del$ can be performed in the Gaussian approximation, which removes $\Del$ from the low-energy spectrum\footnote{Instead of working in the Wilsonian framework of performing the path integral one step at a time, the reader may prefer to study the scalar potential with all fields left in the spectrum. In that case the interaction $\Del^\da Y\Phi\Phi$ of Eq.~(\ref{eq:Delta lagrangian}) induces a small VEV for $\Del^0$, which will have the same effect as Eq.~(\ref{eq:c_1}). In the neutrino literature this is known as the Type-II seesaw mechanism \cite{TypeIIseesaw, TypeIIseesaw2}.} and leaves behind an effective Lagrangian:
\begin{equation}\label{eq:Delta effective lagrangian}
\la_\Del^{\text{eff}} = -\frac{1}{M_\Del^2}\left[ |\A|^2(\ell_i\ell_j)_1(\ell^{\da i}\ell^{\da j})_1+|\B|^2|Y|^2(\Phi_i\Phi_j)_1(\Phi^{\da i}\Phi^{\da j})_1+\left( \A\B\, Y(\Phi^{\da i}\Phi^{\da j})_1(\ell_i\ell_j)_1+h.c.\right)\right]\;.
\end{equation}
The last term in Eq.~(\ref{eq:Delta effective lagrangian}) is an explicit realization of the first term in Eq.~(\ref{eq:dim-5 lagrangian}), with the coefficient
\begin{equation}\label{eq:c_1}
c_1 = \A\B\,\frac{\bra Y\ket}{M_\Del^2}\;.
\end{equation}
This procedure can be repeated for $\Del'$ and $\Del''$ to obtain the second and third operators in Eq.~(\ref{eq:dim-5 lagrangian}) with coefficients
\begin{equation}\label{eq:c_1', c_1''}
c_{1'} = \A'\B' \frac{\bra Y\ket}{M_{\Del'}^2}\;,\;\; c_{1''} = \A''\B'' \frac{\bra Y\ket}{M_{\Del''}^2}\;.
\end{equation}
To leading order there is no term involving $\Del$, $\Del'$, or $\Del''$ that couples to the $A_4$ triplet $(\ell_2\ell_3,\ell_3\ell_1,\ell_1\ell_2)$, so we have $c_3 = 0$. By taking $M_\Del$ large compared to $M_{\Del'}$ and $M_{\Del''}$, or by taking $\A\B$ small compared to $\A'\B'$ and $\A''\B''$, we can make $c_1 \ll c_{1'}, c_{1''}$.
\\\\
Now that we have generated the $c_i$, we need to generate the $\zb_3$-breaking perturbation $\del$. The perturbation away from $\bra\Phi^0\ket \propto (1,1,1)$ can be achieved through a cubic interaction
\begin{equation}
V_{\text{cubic}}(\vec\Phi,\vec S) = \tilde\mu_1\,\vec S\!\cdot\!(\vec\Phi^\da\vec\Phi)_{3_1}+\tilde\mu_2\,\vec S\!\cdot\!(\vec\Phi^\da\vec\Phi)_{3_2}\;.
\end{equation}
If we assume that $\bra\Phi^0_a\ket \equiv \tfrac{1}{\sqrt2}v_a$ are real, then the two terms contribute to the classical potential through one combined invariant: 
\begin{equation}\label{eq:combined invariant}
V_{\text{cubic}}(\vec v,\vec v^{\,S}) = \half\tilde\mu\,(v^S_1v_2v_3+v^S_2v_3v_1+v^S_3v_1v_2)
\end{equation}
where $\tilde\mu \equiv \tilde\mu_1+\tilde\mu_2$. This is the case studied in Appendix~\ref{sec:A4 triplets}. We find that $v_1 = v_3 \neq v_2$ is a solution, with
\begin{equation}\label{eq:splitting}
\del \equiv \half(v_1^2-v_2^2) = \tilde\mu\,\frac{v_S^{(0)}}{\ld_\Phi}+\op(\tilde\mu^2)
\end{equation}
where $v_S^{(0)}$ is the leading order value for $\bra S_2\ket$. Using Eq.~(\ref{eq:Ve3 general form}), we find\footnote{The $c_i$ have mass dimension $-1$, so $V_{e3}$ is dimensionless as required.}
\begin{equation}\label{eq:Ve3 result}
|V_{e3}| \approx \frac{|c_{1'}-c_{1''}|}{|f(x)|}\left| \frac{\tilde\mu v_S^{(0)}}{2\sqrt2\ld_\Phi m_\nu}\right|
\end{equation}
where $f(x) \equiv x/(1-x^2)$ and $x \equiv y_N v_S^{(0)}/m_N$. The various parameters in Eq.~(\ref{eq:Ve3 result}) can be arranged to accommodate the measurement $V_{e3} \sim 0.1$.
\section{Goldstone bosons and dark matter}\label{sec:DM}
Now that we have achieved a realistic neutrino mixing matrix, let us discuss breaking the $U(1)_X$ to a discrete subgroup to avoid generating a massless Goldstone boson after flavor symmetry breaking. Consider the $G_{\text{SM}}\x A_4$-invariant scalar $Y$, which carries charge $-2$ under $U(1)_X$. Instead of the full $U(1)_X$ group, let us instead insist only on the discrete subgroup $Z_4^X: Y \to e^{\,i(+2)2\pi/4}Y = -Y$. Then the most general renormalizable scalar potential for $Y$ is
\begin{equation}\label{eq:V_Y}
V_Y = M_Y^2 Y^\da Y+(\half m_Y^2 Y^2+h.c.)+\ld_Y(Y^\da Y)^2+(\fourth\ld_Y' Y^4+h.c.)\;.
\end{equation}
The terms $Y^2$ and $Y^4$ explicitly break $U(1)_X$ and remove the unwanted Goldstone mode from the low energy spectrum. The $U(1)_X$ representation $\ph \sim m$ for the other fields should now be understood as the transformation property
\begin{equation}\label{eq:Z4}
\zb_4^X: \ph \to e^{\,i(+m)2\pi/4}\ph
\end{equation} so that all fields with $m = 1$ get multiplied by a factor of $i$.
\\\\
Since various aspects of the model [Eq.~(\ref{eq:PH mass relation}) and Eqs.~(\ref{eq:c_1}),~(\ref{eq:c_1', c_1''})] depend crucially on a nonzero $\bra Y\ket$, we take the parameters in Eq.~(\ref{eq:V_Y}) to be such that $\re\,Y$ or $\im\, Y$ (or both) obtains a nonzero VEV. There is no symmetry principle that conserves $Y$-number, so this SM-singlet does not provide a good candidate for dark matter.
\\\\
Given this, one might be motivated to consider a $G_{\text{SM}}\x A_4$-invariant complex scalar $X$ which carries unit charge under $\zb_4^X$:
\begin{equation}\label{eq:X}
X \sim (1,0; 1,+1)\;.
\end{equation}
The scalar potential involving only $X$ and $Y$ is $V(X,Y) = V_X+V_Y+V_{XY}^{\text{cubic}}+V_{XY}^{\text{quartic}}$, where 
\begin{equation}\label{eq:V_X}
V_X = M_X^2 X^\da X+\ld_X(X^\da X)^2+(\fourth \ld_X' X^4+h.c.)\;,
\end{equation}
$V_Y$ was given in Eq.~(\ref{eq:V_Y}), and:
\begin{align}
&V_{XY}^{\text{cubic}} = \left[\half (\rho\, Y+\rho'\, Y^\da)X^2+h.c.\right]\;, \\
&V_{XY}^{\text{quartic}} = \ld_{XY}X^\da X\, Y^\da Y+\left( \half \ld_{XY}' X^\da X\,Y^2+h.c.\right)\;. \label{eq:V_XY^quartic}
\end{align}
If we take $M_X^2$ positive so that $V_X$ does not induce a VEV for $X$, then even when $Y$ picks a VEV and breaks $\zb_4^X$, the potential $V(X,Y)$ leaves a discrete $X$-parity $P_X: X \to -X$ unbroken. Thus, if the $(X,Y)$-sector of the theory were secluded from the rest of the Lagrangian, the lighter of $\re(X)$ and $\im(X)$ would be stable and provide a candidate for dark matter \cite{silveira zee, holz zee}.
\\\\
Now consider the interaction of $X$ with the other fields. In addition to the obvious ``Higgs-portal" interactions such as $X^\da X \Phi^\da\Phi$ which conserve $X$-number, there are quartic interactions which are linear in $X$: 
\begin{equation}\label{eq:V_X^VEV}
V_{\not P_X} = \ka_1 X\phi^{\da i} \vec\Phi_i\! \cdot\! \vec S+\ka_2 X\phi'^{\da i}\e_{ij}\vec\Phi^{\da j}\!\cdot\! \vec S+h.c.
\end{equation}
If we suppose that $\ka_1$ or $\ka_2$ is small but nonzero, then $P_X$ is broken and $X$ can decay. In view of the scales in our model, one might suppose that $M_X \gtrsim$ TeV so that $X$ decays either to the three-body final state $\phi^*\Phi S$ or $\phi'\Phi S$, or if such a decay is kinematically inaccessible, then X can decay to whatever final states can be reached through the propagators of virtual scalars, e.g. TeV-scale neutrinos or SM charged leptons. For the sake of a rough calculation suppose that $M_X \gg M_{\phi'}+M_\Phi+M_S$ and estimate the width (inverse lifetime) of $X$ as
\begin{equation}
\tau_X^{-1} \equiv \G_{X} \sim \frac{\ka^2}{(2\pi)^n}M_X
\end{equation}
where $\ka$ is either $\ka_1$ or $\ka_2$, and $(2\pi)^n$ is a phase space factor for $n$ outgoing particles. Recent measurements of cosmic ray spectra \cite{pamela, fermi} may indicate the presence of late-decaying dark matter, which would require \cite{decaying DM} a lifetime much larger than the age of the universe: $\tau_{\text{min}} \sim 10^{27}$s $\sim 10^{54}$ TeV$^{-1}$, where as usual we work in units $\hbar = 1$. Thus, if we would like to interpret the cosmic ray anomalies in terms of our metastable $X$ particle, we need:
\begin{equation}\label{eq:ratio of lifetimes}
\frac{\tau_X}{\tau_{\text{min}}} \sim (2\pi)^n\left( \frac{\text{TeV}}{M_X}\right)\left( \frac{10^{-27}}{\ka}\right)^{2} \gtrsim 1\;.
\end{equation}
Thus, as is the usual case with a decay rate resulting from renormalizable interactions, we require extremely small coupling constants, $\ka \lesssim 10^{-27}$, if we insist on a DM interpretation of the data. It is worth recalling [see below Eq.~(\ref{eq:V_XY^quartic})] that if we were to impose $P_X: X \to -X$ on the model, then we would have $\ka_1 = \ka_2 = 0$. Taking this argument in the other direction, we note that in the limit $\ka_{1,2} \to 0$ the transformation $P_X: X \to -X$ emerges as a symmetry of the Lagrangian. Thus any value of $\ka$ in Eq.~(\ref{eq:ratio of lifetimes}), no matter how small, is technically natural in the sense of 't Hooft. In order to satisfy Eq.~(\ref{eq:ratio of lifetimes}) we would need to fix $\ka_1$ and $\ka_2$ to be small dimensionless numbers, but we would not need to engineer any unnatural cancellation between large numbers.
\section{Alternative deviation from tribimaximal mixing}\label{sec:alt}
The study of Appendix~\ref{sec:A4 triplets} might lead us to consider the case where $\zb_3 = \{I,c,a\}$ remains a good symmetry of the neutrino mass matrix, while the breaking of $\zb_2 = \{I,r_2\}$ becomes significant. In this case, the neutrino mass matrix in the $G_F$-basis becomes
\begin{equation}
M_\nu = \ml \al&\e_{12}&\beta\\ \e_{12}&\g&\e_{23}\\ \beta&\e_{23}&\al \mr
\end{equation}
with small parameters $\e_{ij}$. If $\U_L^e$ remains approximately unchanged, then $V_{e3} = [(\U_L^e)^\da\U_L^\nu]_{e3} \neq 0$ must arise by modifying the third column of $\U_L^\nu$. In particular,
\begin{equation}
M_\nu \ml \pm 1\\0\\1 \mr = (\al\pm\beta)\ml \pm 1\\0\\1 \mr+(\e_{12}\pm \e_{23})\ml 0\\1\\0 \mr\;.
\end{equation}
Note that if $\e_{12} = +\e_{23}$ then the third column of $\U_L^\nu$ remains unchanged from its leading order value, which is not what we want. This is actually an important observation for our model: the leading order vacuum defined by $\bra\Phi^0\ket \propto (1,1,1)$ and $\bra S\ket \propto (0,1,0)$ leaves invariant a $\zb_2$ symmetry\footnote{This $\zb_2$ operation does not commute with $A_4$. The reason this phenomenon occurs is that a scalar potential which involves only an $A_4$ triplet of $SU(2)\x U(1)$ doublets and an $A_4$ triplet of $SU(2)\x U(1)$ singlets is accidentally invariant under the permutations with $\det = -1$ in addition to the cyclic and anticyclic permutations ($\det = +1$) which are elements of $A_4$. The charged lepton Yukawa interactions are not invariant under the odd permutations, and so the theory will in general be invariant only under the smaller group $A_4$ when higher order corrections are included.} which interchanges $1\lrarrow3$ [see Eqs.~(\ref{eq:perturbed S vacuum}) and~(\ref{eq:perturbed phi vacuum})]. Thus any tree-level perturbation from the scalar potential will result in $\e_{12} = \e_{23}$, and so this simplest approach will not generate a nonzero $V_{e3}$.
\\\\
What we need instead is to perturb around the leading order vacuum $\bra\Phi^0\ket \propto (1,1,-1)$ and $\bra S\ket \propto (0,1,0)$, which will leave behind a (tree-level) symmetry which exchanges $1\lrarrow3$ with an additional minus sign \cite{A4ma2}. In this case there is no clash between $\zb_3$ and $\zb_2$, since the vacuum for $\Phi$ breaks the $\zb_3$ explicitly; instead there is a clash between the two sectors simply because $\bra S\ket$ preserves $\zb_2$ while $\bra\Phi\ket$ does not. In this case the charged lepton Yukawa matrix in the $G_F$-basis becomes
\begin{equation}\label{eq:new yukawa matrix}
Y_e = \ml -y_e & -y_e' & -y_e''\\ y_e & \w\,y_e' &\w^*y_e''\\ y_e & \w^*y_e' & \w\,y_e'' \mr\;.
\end{equation}
The left-diagonalization matrix\footnote{In Eq.~(\ref{eq:modified diagonalization}) we drop the phase angles $a_e,b_e,c_e$ for notational clarity.} of Eq.~(\ref{eq:U_L^f}) is now modified by a factor of $(-1)$ in its first row, such that its transpose diagonalizes $M_e$:
\begin{equation}\label{eq:modified diagonalization}
(\U_L^e)^TM_e \propto \ml -1&1&1\\ -1&\w^*&\w\\ -1&\w&\w^* \mr \ml -y_e & -y_e' & -y_e''\\ y_e & \w\,y_e' &\w^*y_e''\\ y_e & \w^*y_e' & \w\,y_e'' \mr \propto \text{diag}(y_e, y_e',y_e'')\;.
\end{equation}
Before we take into account the tree-level vacuum perturbations due to the interactions between $\Phi$ and $S$, we still have the same neutrino mass matrix of Eq.~(\ref{eq:seesaw mass matrix}). Due to the extra signs in Eq.~(\ref{eq:new yukawa matrix}) relative to Eq.~(\ref{eq:U_L^f}), the leading order tribimaximal mixing matrix $V = V_{\text{TB}}$ is obtained by switching the first and third columns of Eq.~(\ref{eq:U_L^nu})\footnote{By the absolute value notation in Eq.~(\ref{eq:VTB again}) we mean simply to take the absolute value of each entry in $V$ after performing the matrix multiplication.}:
\begin{equation}\label{eq:VTB again}
|V| \equiv \left| (\U_L^e)^\da \U_L^\nu \right| \propto \left| \ml -1&1&1\\ -1&\w&\w^*\\ -1&\w^*&\w \mr \ml -1&0&+1\\ 0&\sqrt2&0\\ +1&0&+1 \mr\right| = \ml 2&\sqrt2 &0\\ 1&\sqrt2&\sqrt3\\ 1& \sqrt2 &\sqrt3 \mr\;.
\end{equation}
The operation of permuting the first and third columns of $\U_L^\nu$ simply permutes the first and third masses\footnote{Let the 3-by-3 complex symmetric matrix $M$ be Takagi-diagonalized by a 3-by-3 unitary matrix $U$: $U^T M U = D$ with $D$ a diagonal matrix whose nonzero entries are real and positive. Let $C$ be a permutation matrix, and define $\tilde U \equiv UC$, which is the matrix $U$ with its columns permuted. Then we find $\tilde U^T M \tilde U = CDC \equiv \tilde D$, where the matrix $\tilde D$ is simply the matrix $D$ with its diagonal entries permuted according to the permutation defined by $C$.} in Eq.~(\ref{eq:neutrino masses}), which can be compensated for by taking the opposite sign of the parameter $x$.
\\\\
Using the results of Appendix~\ref{sec:A4 triplets} we find a perturbation with $\e_{12} = -\e_{23}$, which results in a nonzero $V_{e3}$:
\begin{align}
&|V_{e3}| \approx \e \sqrt{\frac{2}{3}}\left| \frac{1-x}{x}\right| + \op(\e^2)\;.
\end{align}
Here $\e \equiv |\e_{12}| = |\e_{23}| \sim (v/v_S)^2 (\tilde\mu/v_S)$ [see Eq.~(\ref{eq:alt perturbation to (1,0,0)})], and $\tilde\mu$ was defined in Eq.~(\ref{eq:combined invariant}). The benefit of this alternative derivation of a nonzero reactor angle is that it is easier to reconcile with the best-fit value for the solar angle, and it requires no additional scalar fields (e.g. $SU(2)$ Higgs triplets) beyond those which we have already introduced to obtain the leading order expressions for $M_e$ and $M_\nu$. In order to ignore the perturbations to $\bra\Phi\ket \propto (1,1,-1)$ induced by the interaction with $S$, we must take $v_S \lesssim v$. Thus an additional requirement of this approach is that the SM-singlet scalars $S_a$ have masses bounded from above by $\sim 100$ GeV (under the assumption that perturbation theory remains valid). There is, however, no additional symmetry which prevents their prompt decay into lighter particles, so they are not stable enough to constitute the majority of dark matter.
\section{Discussion}\label{sec:end}
We have embedded the HKV model into a framework whose largest scale $\M$ is much smaller than the traditional seesaw scale: $\M/M_{\text{GUT}} \sim 10^{-10}$. The main requirements of such a scenario are an additional $A_4$-singlet Higgs doublet, $\phi'$, and an SM-singlet complex scalar, $Y$, which carry $U(1)_X$ charge $+1$ and $-2$, respectively [Eqs.~(\ref{eq:Y}) and~(\ref{eq:phi'})]. Then the scalar potential can be arranged such that the small Dirac mass for the neutrino can be generated by a Higgs doublet whose VEV is small but whose mass is large [Eq.~(\ref{eq:PH mass relation})]. 
\\\\
This ameliorates the sequestering problem that typically afflicts models of this type for the simple reason that the separation between the scale at which $G_F = A_4\x U(1)_X$ is imposed and the scale of neutrino masses is not sufficiently vast to induce large renormalization group corrections. This is an alternative to the supersymmetric and brane-world approaches mentioned by HKV and studied by various authors.
\\\\
The recent measurement of $V_{e3} \sim 0.1$ is accommodated by a minimal modification to tribimaximal mixing due to $A_4$-singlet Higgs triplets which communicate the breaking of $\zb_3 = \{I,c,a\}$ into the neutrino sector while leaving the $\zb_2 = \{I,r_2\}$ essentially unbroken (see Section~\ref{sec:Ve3}). An alternative possibility is to perturb the theory around the vacuum $\bra\Phi^0\ket \propto (1,1,-1)$ and feed the corrections due to breaking the $\zb_2 = \{I,r_2\}$ into the neutrino mass matrix (see Section~\ref{sec:alt}).
\\\\
The desire to remove an unwanted Goldstone boson due to breaking the continuous global $U(1)_X$ led us to insist only on the discrete subgroup $\zb_4^X < U(1)_X$, and the Lagrangian for the SM-singlet $Y$ explicitly implements the distinction between the two symmetry groups [Eq.~(\ref{eq:V_Y})]. This then motivated (but did not require) the introduction of a second SM-singlet complex scalar field $X$ with half the $\zb_4^X$ charge of $Y$, which can be a candidate for decaying dark matter [Eqs.~(\ref{eq:X}) and~(\ref{eq:V_X^VEV})].
\\\\
We conclude with a brief evaluation of the relative theoretical merits of the ``complicated" theory of Section~\ref{sec:Ve3} as compared to the ``simple" theory of Section~\ref{sec:alt}. On the one hand, it is often desirable for computational simplicity to assume a principle of economy in low-energy phenomenology, in which case one might prefer not to add any additional fields besides those that generate the leading order results for $M_e$ and $M_\nu$. On the other hand, it is well-known that many attempts to provide a high-energy completion for the SM [e.g. embedding $SU(3)\x SU(2)\x U(1)$ into a larger group such as $SO(10)$] require a large number of scalar fields. It would be an exciting possibility if these additional fields contribute observable effects to low-energy phenomenology such as neutrino mixing.
\\\\
\textit{Acknowledgments:}
\\\\
This research was supported by the NSF under Grant No. PHY07-57035.
\appendix
\section{$A_4$ invariant potential for a real triplet}\label{sec:A4 triplet}
The scalar potential for a Higgs field, $\ph$, transforming as an $A_4$ triplet was studied previously by one of the authors \cite{zeeA4}. Invariance under $SU(2)\x U(1)$ implies that this potential is a function only of the combination $\ph^\da_a\ph_b$, where $a,b$ label the $A_4$ components, and it has one quadratic invariant and five quartic invariants.
\\\\
Here we study the scalar potential for a real scalar field $S$ transforming as a triplet under $A_4$. Since $S$ is real, the five quartic invariants for $\ph$ collapse to only two independent terms. However, since we do not impose a parity that flips the sign of $S$ (the real-valued analog of a $U(1)$ symmetry), there is also a nontrivial cubic invariant from the group theoretic property $3\x3\x3 = 3\x(1+1'+1''+3_A+3_S)$ = $3\x3_S+... = 1+...$, where the 1 denotes the invariant $S_1S_2S_3$. (Note that $3_A = 0$ since $S_aS_b = S_bS_a$.)
\\\\
The most general renormalizable, $A_4$-invariant scalar potential for $S \sim 3$ is
\begin{align}
V(S) &= \half M^2 \sum_{a\,=\,1}^3 S_a^2+\mu S_1S_2S_3+\fourth\ld\left( \sum_{a\,=\,1}^3 S_a^2\right)^2+\half\ld'\left[ (S_1S_2)^2+(S_2S_3)^2+(S_3S_1)^2\right]\;.
\end{align}
We first study the three potential vacua that were discussed in \cite{zeeA4}: \\\\
$E: \{\bra S_1\ket = \bra S_2\ket = \bra S_3\ket \equiv S \neq 0\}$, $U: \{\bra S_1\ket \equiv S \neq 0,\, \bra S_2\ket = \bra S_3\ket = 0\}$, and $P: \{\bra S_1\ket = \bra S_2\ket \equiv S \neq 0,\, \bra S_3\ket = 0\}$.
\\\\
In the vacuum $E$ we have $M^2 = -(3\ld+2\ld')S^2-\mu S$, and:
\begin{align}
&\left. V\right|_E = -\tfrac{9}{4}(\ld+\tfrac{2}{3}\ld')S^4-\half\mu S^3\qquad\text{ and }\qquad \left. \frac{\pa^2V}{\pa S_a \pa S_b}\right|_E = \ml m^2&\Del&\Del\\ \Del&m^2&\Del\\ \Del&\Del&m^2 \mr
\end{align}
where $m^2 \equiv 2\ld S^2-\mu S\;,\;\;\Del \equiv 2(\ld+\ld')S^2+\mu S$. The notation $\left. V\right|_{E}$ denotes the value of the potential evaluated at the minimum for the vacuum $E$. Diagonalizing the matrix of second derivatives gives $\D_E^2 = \text{diag}(m^2+2\Del,m^2-\Del,m^2-\Del)$. The requirement $m^2-\Del \geq 0$ implies $\mu S \leq -\half \ld' S^2$. 
\\\\
In the vacuum $U$ we have $M^2 = -\ld S^2$, and:
\begin{align}
&\left. V\right|_U = -\fourth\ld S^4\qquad\text{ and }\qquad \left. \frac{\pa^2V}{\pa S_a \pa S_b}\right|_U = \ml  m_0^2&0&0\\0&m^2&\Del\\0&\Del&m^2 \mr
\end{align}
where $m_0^2 \equiv 2\ld S^2$, $m^2 \equiv (\ld'-\ld)S^2$, and $\Del \equiv \mu S$. The matrix of mass-squared eigenvalues is $\D^2_U = \text{diag}(m_0^2,m^2-2\Del,m^2+2\Del)$. The requirement $m^2-2\Del \geq 0$ is satisfied if $\mu S \leq \half (\ld'-\ld)S^2$.   
\\\\
When studying the putative vacuum $P$ we find an important difference from the previously studied Higgs potential. Minimizing the potential with respect to $S_3$ results in the equation
\begin{equation}
\left. \frac{\pa V}{\pa S_3}\right|_P = \mu S^2 = 0\;.
\end{equation}
If $\mu \neq 0$ then this equation cannot be satisfied. Thus, unlike the case of a complex $SU(2)\x U(1)$-invariant Higgs potential, a real scalar triplet does not admit the vacuum $P$ even as a local minimum.\footnote{In the $SU(2)\x U(1)$ case $P$ is a local minimum but never the global one \cite{zeeA4}.}
\\\\
As pointed out by E. Ma \cite{A4ma2}, it is worth also studying the cases $E': \{ \bra S_1\ket = \bra S_2\ket = -\bra S_3\ket \equiv S \neq 0 \}$ and $P': \{ \bra S_1\ket = -\bra S_2\ket \equiv S\neq 0\;,\;\; \bra S_3\ket = 0 \}$. The sign flip simply changes the effective sign of $\mu$, since the other terms in the potential are all quadratic in each $S_a$. This is immaterial for the simple case studied in this section, but it will be important when we study the potential for two interacting $A_4$ triplets.
\section{Interactions between a Higgs field and a real triplet}\label{sec:A4 triplets}
Now we study in detail whether the VEVs $\bra S\ket \propto (1,0,0)$ and $\bra\ph\ket \propto (1,1,1)$ can hold to a good approximation in a Lagrangian based on $A_4\x \zb_2$. Let $\ph \sim 3_-$ and $S \sim 3_+$ be two real triplets of $A_4$, where the subscript indicates the parity under $\zb_2$. The real field $\ph$ which is odd under $\zb_2$ serves as a toy model for a complex field $\Phi$ which is charged under $SU(2)\x U(1)$. 
\\\\
The potential is $V = V_S+V_\ph+V_{S\ph}^{\text{cubic}}+V_{S\ph}^{\text{quartic}}$, where
\begin{align}
&V_S = \half M_S^2 \sum_{a\,=\,1}^3 S_a^2+\mu_S S_1S_2S_3+\fourth\ld_S\left( \sum_{a\,=\,1}^3 S_a^2\right)^2+\half\ld_S'\left[ (S_1S_2)^2+(S_2S_3)^2+(S_3S_1)^2\right] \label{eq:V_S}\\
&V_\ph =  \half M_\ph^2 \sum_{a\,=\,1}^3 \ph_a^2+\fourth\ld_\ph\left( \sum_{a\,=\,1}^3 \ph_a^2\right)^2+\half\ld_\ph'\left[ (\ph_1\ph_2)^2+(\ph_2\ph_3)^2+(\ph_3\ph_1)^2\right] \label{eq:V_phi}
\end{align}
describes each sector, and
\begin{align}
&V_{S\ph}^{\text{cubic}} = \tilde\mu(S_1\ph_2\ph_3+S_2\ph_3\ph_1+S_3\ph_1\ph_2) \label{eq:V_Sphi^cubic}\\
&V_{S\ph}^{\text{quartic}}\! =\!\ka_0(\sum_a \!S_a^2)(\sum_b\! \ph_b^2)\!+\!\ka_1(S_1^2\ph_2^2\!+\!\text{cyclic})\!+\!\ka_2(\ph_1^2S_2^2\!+\!\text{cyclic})\!+\!\ka_3(S_1S_2\ph_1\ph_2\!+\!\text{cyclic}) \label{eq:V_Sphi^quartic}
\end{align}
describe the interactions between the two sectors.
\\\\
We would like to find a perturbative solution around the configurations $S \propto (1,0,0)$ and $\ph \propto (1,1,1)$. The general case can be parametrized as
\begin{equation}
S_a = S_a^{(0)}+\e S_a^{(1)}\;,\;\; \ph_a = \ph_a^{(0)}+\e \ph_a^{(1)}
\end{equation}
where $\e$ is a small parameter. We will first assume that $V_{S\ph}^{\text{quartic}}$ can be dropped, or in other words that the $\ka_i$ are at most $\op(\e^2)$ and thus negligible. We will assume that $V_{S\ph}^{\text{cubic}} \equiv \e V_1$ is itself a perturbation by defining $\tilde\mu = \e\tilde\mu_0$ with $\tilde\mu_0$ a parameter of $\op(\mu_S)$. The potential $V_0 \equiv V_S+V_\ph$ is to be treated as the leading order potential of $\op(\e^0)$, with any possible renormalization group corrections to its parameters being of $\op(\e^2)$ or higher. 
\\\\
The goal is to solve the equations $\frac{\pa}{\pa S_a}(V_0+\e V_1) = 0$ and $\frac{\pa}{\pa\ph_a}(V_0+\e V_1) = 0$ to $\op(\e)$. In the particular leading order vacuum of interest, $S_a^{(0)} \propto (1,0,0)$ and $\ph_a^{(0)} \propto (1,1,1)$, this procedure gives the result
\begin{align}
&S_a = \ml 1\\0\\0 \mr_a S_0+\e S_a^{(1)}\;,\;\; \ph_a = \ml 1\\ 1\\ 1 \mr_a \ph_0+\e \ph_a^{(1)}
\end{align}
where:
\begin{align}
&S_1^{(1)} = -\,\frac{1}{2\ld_S}\left( \frac{\ph_0}{S_0}\right)^2 \tilde\mu_0\;,\;\; S_2^{(1)} = S_3^{(1)} = -\,\frac{1}{\ld_S'+\frac{\mu_S}{S_0}}\left( \frac{\ph_0}{S_0}\right)^2\tilde\mu_0\;, \label{eq:perturbed S vacuum}\\
&\nonumber\\
&\ph_1^{(1)} = -\,\frac{\ld_\ph+\ld_\ph'}{\ld_\ph(\ld_\ph+\tfrac{3}{2}\ld_\ph')}\left( \frac{S_0}{\ph_0}\right)\tilde\mu_0\;,\;\; \ph_2^{(1)} = \ph_3^{(1)} = +\,\frac{\ld_\ph'}{2\ld_\ph(\ld_\ph+\tfrac{3}{2}\ld_\ph')}\left( \frac{S_0}{\ph_0}\right)\tilde\mu_0\;. \label{eq:perturbed phi vacuum}
\end{align}
If we take $\ph_0/S_0 \lesssim 1$, then the corrections to $S_a \propto (1,0,0)$ can be ignored safely, but those to $\ph_a \propto (1,1,1)$ are important. This is precisely the situation we seek in the main text: the $\zb_3$ ``charged lepton" subgroup of $A_4$ is broken, but the $\zb_2$ ``neutrino" subgroup of $A_4$ remains conserved to a good approximation. The effect we are talking about is of order $S_2^{(1)}/(\ph_2^{(1)}-\ph_1^{(1)}) \sim (\ph_0/S_0)^3$, e.g. $\sim (1/4)^3 \sim 1\%$.
\\\\
A convenient way to parametrize this breaking of $\zb_3 < A_4$ is:
\begin{equation}
\del \equiv \half(\ph_2^{\,2}-\ph_1^{\,2}) = \e\tilde\mu_0\,\frac{S_0}{\ld_\ph}+\op(\e^2)\;.
\end{equation}
This is the equation we use to obtain Eq.~(\ref{eq:splitting}) in the main text.
\\\\
Notice that the corrections of Eqs.~(\ref{eq:perturbed S vacuum}) and~(\ref{eq:perturbed phi vacuum}) respect the symmetry $2\lrarrow3$. In order to find a perturbation that does not respect this symmetry, we can break it explicitly by choosing the leading order vacuum $\ph_a \propto (1,1,-1)$ and $S_a \propto (1,0,0)$. Carrying out the same procedure as above gives
\begin{align}\label{eq:alt perturbation to (1,0,0)}
& S_1^{(1)} = +\frac{1}{2\ld_S'}\left( \frac{\ph_0}{S_0}\right)^2\tilde\mu_0\;,\;\; S_2^{(1)} = -S_3^{(1)} = +\,\frac{1}{\ld_S-\frac{\mu_S}{S_0}}\left(\frac{\ph_0}{S_0}\right)^2\tilde\mu_0
\end{align}
for the corrections to $S_a^{(0)} = S_0(1,0,0)$, and
\begin{align}
& \ph_1^{(1)} = +\left\{\frac{\ld_\ph+\ld_\ph'}{(2\ld_\ph+3\ld_\ph')\ld_\ph+4(\ld_\ph')^2}\right\}\,\frac{\ld_\ph'}{\ld_\ph}\left( \frac{S_0}{\ph_0}\right)\tilde\mu_0\\
&\ph_2^{(1)} = -\frac{1}{2} \left\{ \frac{(2\ld_\ph+3\ld_\ph')\ld_\ph+2(\ld_\ph')^2}{\ld_\ph\left[(2\ld_\ph+3\ld_\ph')\ld_\ph+4(\ld_\ph')^2 \right]} \right\}\left( \frac{S_0}{\ph_0}\right)\tilde\mu_0\\
&\ph_3^{(1)} = +\frac{1}{2}\left\{ \frac{2\ld_\ph+3\ld_\ph'}{(2\ld_\ph+3\ld_\ph')\ld_\ph+4(\ld_\ph')^2} \right\}\left( \frac{S_0}{\ph_0}\right)\tilde\mu_0
\end{align}
for the corrections to $\ph_a^{(0)} = \ph_0(1,1,-1)$. If we take $S_0 \lesssim \ph_0$ then only the corrections to $S_a$ are important, and we generate a perturbation to the vacuum $S_a \propto (1,0,0)$ which is antisymmetric in the exchange $2\lrarrow3$.
\section{RG for toy model coupled to fermions}\label{sec:RG}
\begin{figure}[h]
\begin{center}
\fbox{
	\begin{minipage}{18 cm}
		\subfigure[$g(\mu)$]
		{
		\includegraphics[scale=0.85]{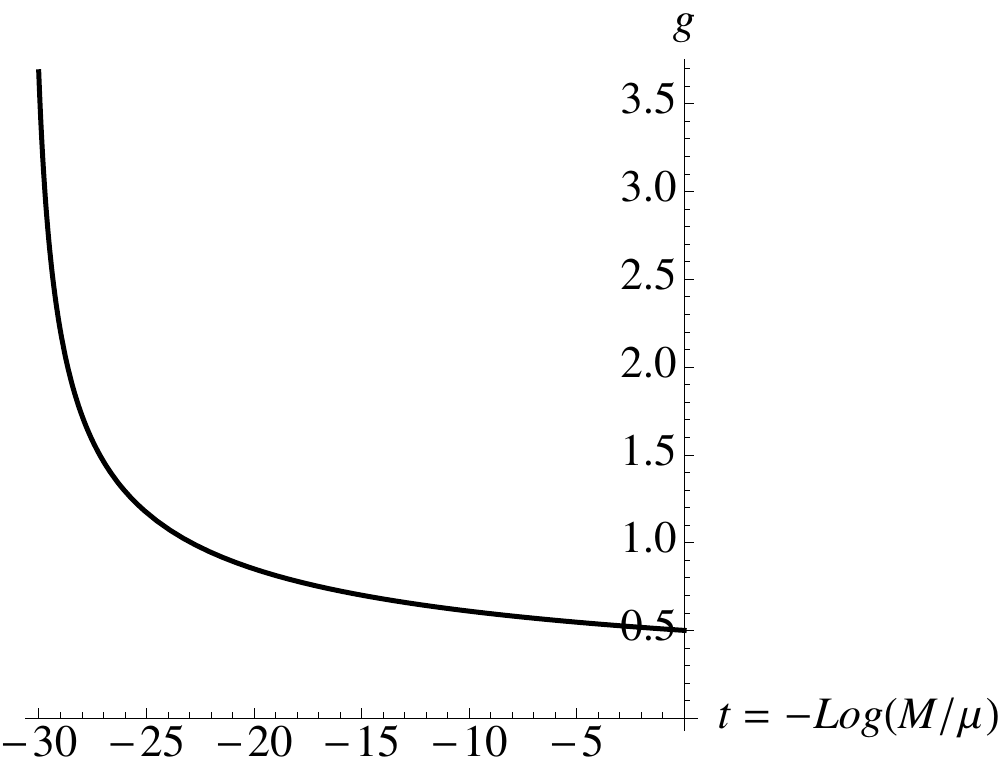}
		\label{fig:g}
		}
		\subfigure[$y(\mu)$]
		{
		\includegraphics[scale=0.85]{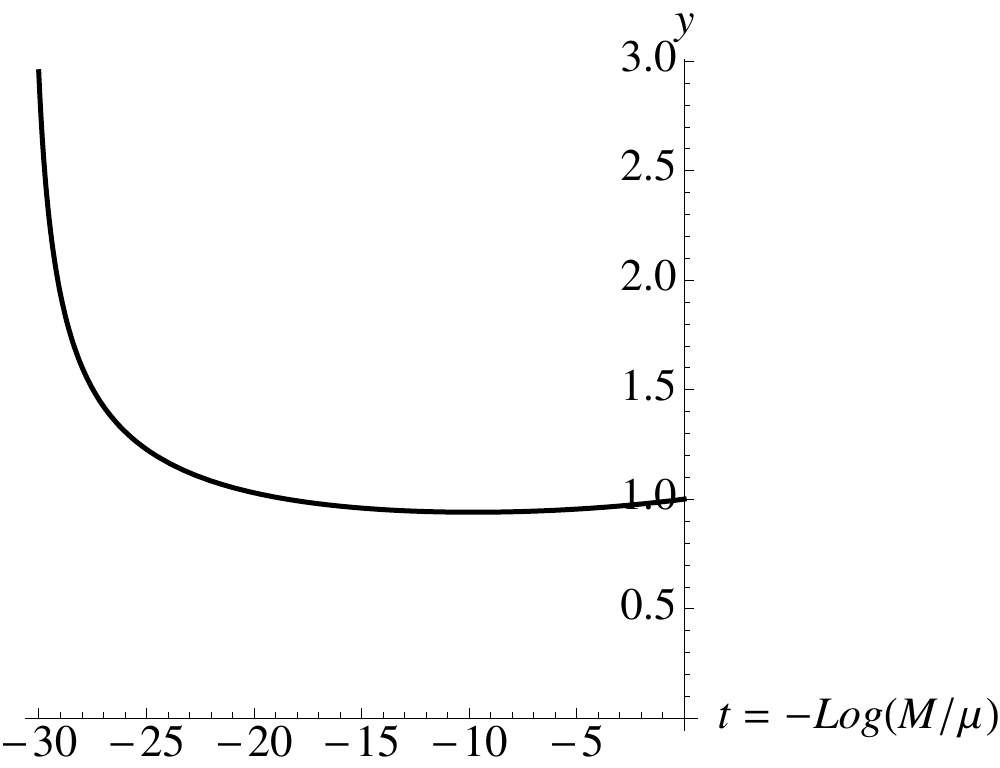}
		\label{fig:y}
		}
		\vspace{-10pt}
		\caption{\small{The running gauge coupling $g(\mu)$ and Yukawa coupling $y(\mu)$ in the toy model of Eq.~(\ref{eq:toy lagrangian}).} }
	\end{minipage}
	}
\end{center}
\end{figure}

\begin{figure}[h]
\begin{center}
\fbox{
	\begin{minipage}{18 cm}
		\subfigure[$\ld_1(\mu)$]
		{
		\includegraphics[scale=0.85]{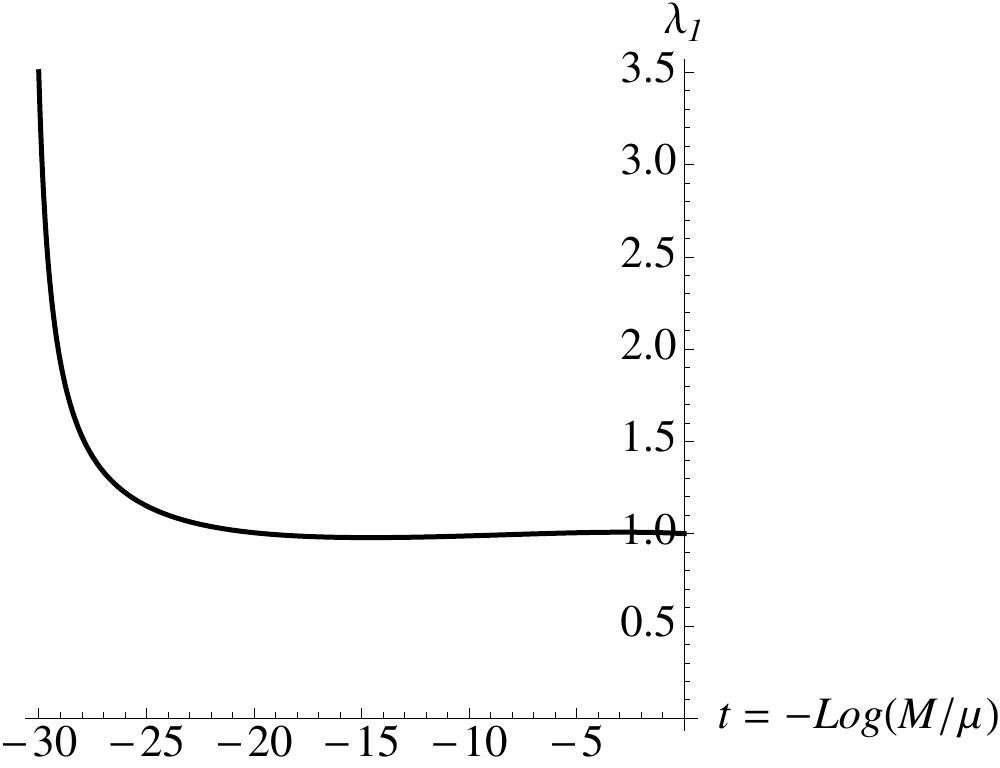}
		\label{fig:ld1}
		}
		\subfigure[$\ld_2(\mu)$]
		{
		\includegraphics[scale=0.85]{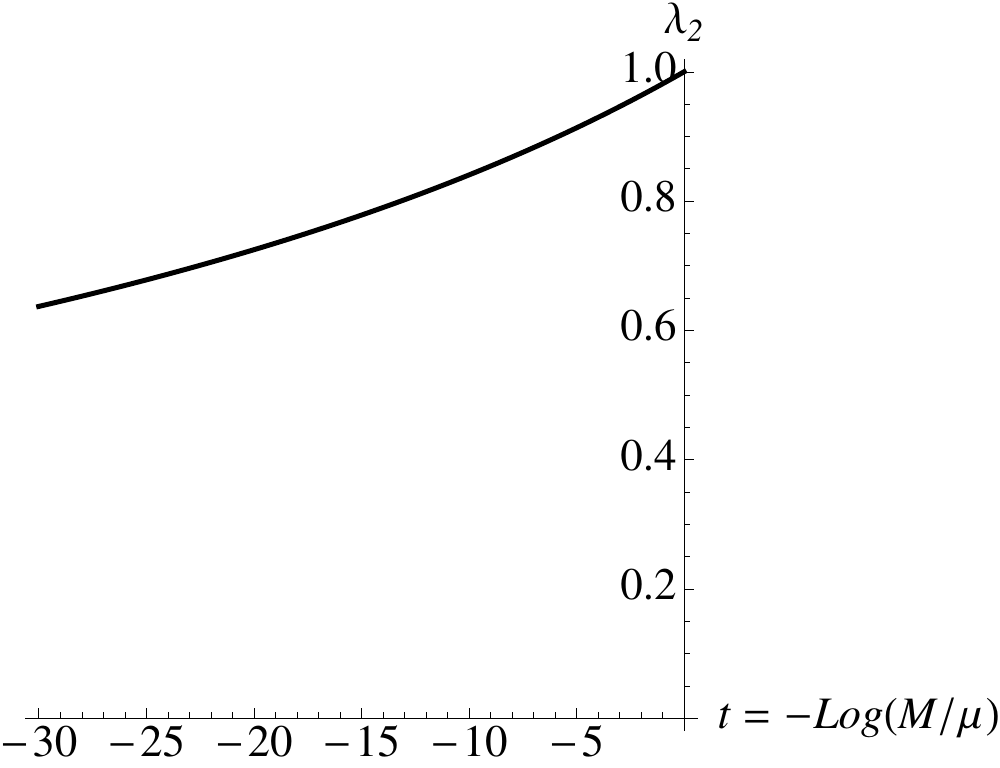}
		\label{fig:ld2}
		}
		\vspace{-10pt}
		\caption{\small{The running quartic self-couplings $\ld_1(\mu)$ and $\ld_2(\mu)$ in the toy model of Eq.~(\ref{eq:toy lagrangian}).} }
	\end{minipage}
	}
\end{center}
\end{figure}

\begin{figure}[h]
\begin{center}
\fbox{
	\begin{minipage}{14 cm}
	\begin{center}
		\includegraphics[scale=1]{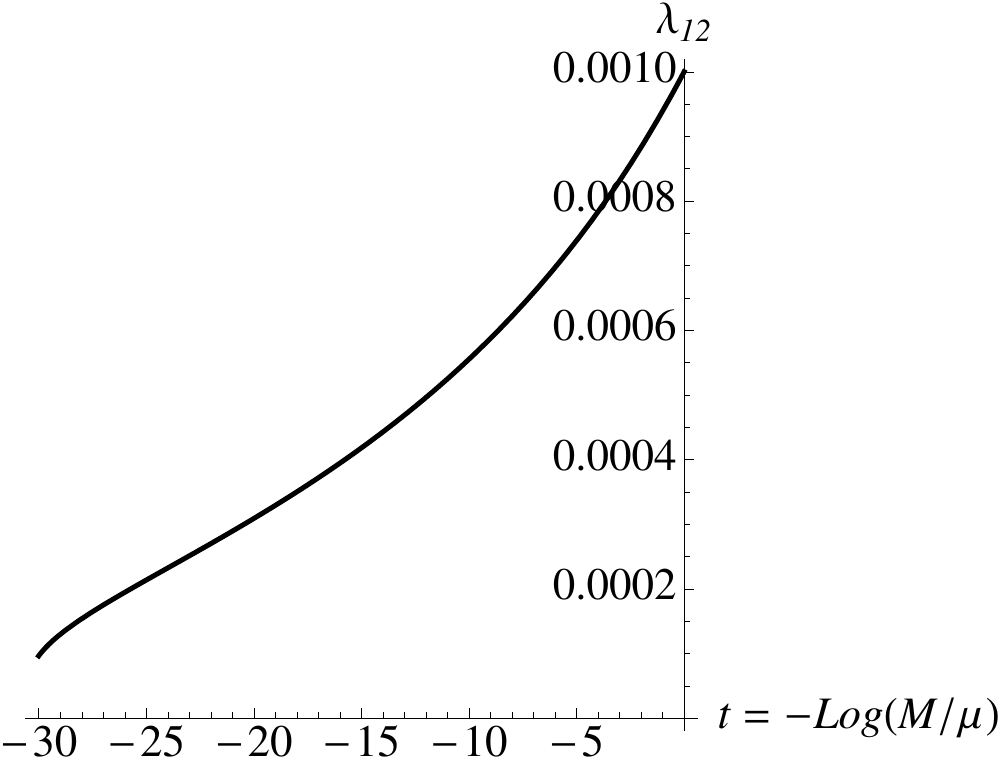}
	\end{center}
	\vspace{-10pt}
	\caption{\small{The running quartic coupling $\ld_{12}(\mu)$ which couples $\phi_1$ and $\phi_2$ in the toy model of Eq.~(\ref{eq:toy lagrangian}).}}
	\label{fig:ld12}	
	\end{minipage}
	}
\end{center}
\end{figure}

Consider a simple toy model of two real scalars $\phi_1,\phi_2$ coupled to ``quarks" $q$ and ``antiquarks" $\bar q$, with a color gauge group $SU(N)$ under which $q$ and $\bar q$ transform as $N$ and $N^*$, respectively. (The $\phi_i$ are color-neutral, and the gluons are denoted by $\{G_\mu^a\}_{a\,=\,1}^{N^2-1}$.) We impose the reflection symmetry\footnote{This $\zb_2$ serves as a toy model for $SU(2)$ gauge symmetry, under which Higgs fields (``$\phi_1$") and left-handed quarks transform, but under which SM-singlet scalars (``$\phi_2$") and left-handed antiquarks are neutral.}
\begin{equation}\label{eq:Z2 reflection}
\zb_2:\qquad \phi_1 \to -\phi_1,\;q \to -q,\; \phi_2 \to +\phi_2, \;\bar q \to +\bar q
\end{equation}
so that $\phi_1$ has a Yukawa interaction with $q\bar q$ whereas $\phi_2$ does not:
\begin{equation}\label{eq:toy yukawa}
\la_{\text{Yuk}} = -y\phi_1q\bar q+h.c.
\end{equation}
The most general renormalizable potential for $\phi_1$ and $\phi_2$ consistent with $\zb_2$ symmetry is:
\begin{equation}\label{eq:toy potential}
V(\phi_1,\phi_2) = \half M_1^2\phi_1^2+\half M_2^2\phi_2^2+M_{12}\,\phi_1^2\phi_2+\tfrac{1}{4!}\ld_1\phi_1^4+\tfrac{1}{4!}\ld_2\phi_2^4+\tfrac{1}{4}\ld_{12}\phi_1^2\phi_2^2\;.
\end{equation}
We are interested in the renormalization group equations (RGE) for the Lagrangian
\begin{equation}\label{eq:toy lagrangian}
\la_{\text{toy}}(G,\phi_1,\phi_2,q,\bar q) = \la_{\text{kin}}(G,\phi_1,\phi_2,q,\bar q)+\la_{\text{Yuk}}(\phi_1,q,\bar q)-V(\phi_1,\phi_2)
\end{equation}
in the deep UV regime, far above the scalar masses $M_1, M_2$ and the dimensionful cubic coupling $M_{12}$. (Since we are not interested in low-energy phenomena such as spontaneous symmetry breaking, we take $M_1^2$ and $M_2^2$ both positive.)
\\\\
The issue we would like to address is the following. Let $m$ denote a ``low-energy" mass scale $m \sim M_1 \sim M_2 \sim M_{12}$, and let $\M$ denote the UV cutoff of the field theory. Suppose we (arbitrarily) fix the parameters of the theory at the scale $\M$ to satisfy 
\begin{equation}
\ld_1(\M) \sim \ld_2(\M) \sim O(1)\;,\;\; |\ld_{12}(\M)| \ll  1\;.
\end{equation}
Our goal is then to find the maximum ratio $\M/m$ such that the ``off-diagonal" coupling $\ld_{12}$ remains parametrically smaller than the ``diagonal" couplings $\ld_1$ and $\ld_2$ at energy $\mu \sim m \ll \M$.
\\\\
This toy model is a simple subcase of a general class of theories studied by Cheng, Eichten, and Li \cite{RG1}.  Following their notation, we express our Lagrangian as\footnote{In contrast to the more general case studied in the reference, but similarly to the realistic case in the SM and many of its extensions, here our scalars $\phi_i$ are not charged under the color group. Also here we need only one flavor of quark, in which case $h_i$ is a single number (for each $i$) rather than a matrix.}
\begin{align}
&\la = -\fourth G^a_{\mu\nu} G^{a\mu\nu}\!\!-\!\half \pa_\mu\phi_i\pa^\mu\phi_i\!+\! iq^\da \bar{\not \!\!D} q\!+\! i\bar q^\da\bar {\not \!\! D}\bar q\!-\!(q\,h_i\bar q\,\phi_i+h.c.)\!-\!V(\phi)\;, \\
&V(\phi) = \tfrac{1}{4!}f_{ijk\ell} \phi_i\phi_j\phi_k\phi_{\ell} + \text{lower-dimension terms}\;.
\end{align}
Here $G^a_{\mu\nu} = \pa_\mu G^a_\nu-\pa_\nu G^a_\mu+gC^{abc}G^b_\mu G^c_\nu$ is the gluon field strength, and $\bar {\not\!\! D} = \bar\s^\mu D_\mu$ is the covariant derivative for each 2-component fermion: $D_\mu q = (\pa_\mu-igG_\mu^at^a)q$ and $D_\mu \bar q = (\pa_\mu-igG_\mu^a\bar t^a)\bar q$ with $\bar t^a = -(t^a)^*$.
\\\\
The lowest-order\footnote{As pointed out in the reference, the last term of Eq.~(\ref{eq:beta function h}) is of ``lowest-order" even though it arises from a 2-loop diagram.} RGE for the couplings $g, h_i$, and $f_{ijk\ell}$ are $(\D \equiv 16\pi^2d/dt,\,t \equiv \ln\frac{\mu}{\mu_0}$ for the scale $\mu$ and an arbitrary reference point $\mu_0$):
\begin{align}
&\mathcal Dg = -\half b_0 g^3\;,\;\; b_0 = 2\left(\tfrac{11}{3}S_1^{\mathcal G}-\tfrac{4}{3}S_3^{F}\right) \label{eq:beta function g}\\
&\nonumber\\
&\mathcal Dh_i\! =\! 2h_m h_i h_m\!+\!\half(h_m h_m h_i\!+\! h_i h_m h_m)\!+\!2\tr(h_i h_m) h_m \nonumber\\
&\qquad-\!3(2t^a h_i t^a\!+\!S_2^{F}h_i)g^2\!+\!\tfrac{1}{288\pi^2} f_{ijk\ell} f_{jk\ell m}h_m \label{eq:beta function h}\\
&\nonumber\\
&\D f_{ijk\ell}\!=\! f_{ijmn}f_{mnk\ell}\!+\!f_{ikmn}f_{mnj\ell}\!+\!f_{i\ell mn}f_{mnjk}\!+\!8\tr(h_ih_m)f_{mjk\ell} \nonumber\\
&\qquad\!-\!2\tr(h_ih_j\{h_k,h_\ell\}\!+\!h_ih_k\{h_j,h_\ell\}\!+\!h_i h_\ell \{h_j,h_k\}) \label{eq:beta function f}
\end{align}
Here we have defined the group theoretic quantities $C^{acd}C^{bcd} = S_1^{\mathcal G}\del^{ab}$, $t^a t^a = S_2^FI$, and $\tr(t^a t^b) = S_3^F\del^{ab}$ related to the color representation of the fermions\footnote{Our two-component quarks are coupled in a ``vectorlike" fashion to the gluons, so it is convenient to package them into 4-component Dirac spinors $Q = (q,\bar q^\da)$, which here transform under the defining $N$-dimensional representation of $SU(N)$.] }. For $SU(N)$, we have $S_1^{\mathcal G} = N$, $S_2^F = \tfrac{N^2-1}{2N}$, and $S_3^F = \half$.
\\\\
For our case [Eqs.~(\ref{eq:toy yukawa}) and~(\ref{eq:toy potential})] we have:
\begin{align}
&h_i = y\,\del_{i1}\;,\\
&f_{1111} = \ld_1\;,\;\; f_{2222} = \ld_2\;,\\
&f_{1122} = f_{1212} = f_{1221} = f_{2121} = f_{2112} = f_{2211} = \ld_{12}\;,\\
&f_{1112} = f_{1121} = f_{1211} = f_{2111} = f_{2221} = f_{2212} = f_{2122} = f_{2221} = 0\;.
\end{align}
Then Eq.~(\ref{eq:beta function h}) simplifies to
\begin{equation}\label{eq:beta function y}
\D y= 5y^3-9S_2^F y g^2+\tfrac{1}{288\pi^2}(\ld_1^2+3\ld_{12}^2)y
\end{equation}
and Eq.~(\ref{eq:beta function f}) produces three reasonably simple coupled equations:
\begin{align}
&\D \ld_1 = 3(\ld_1^2+\ld_{12}^2)+8y^2\ld_1-12y^4 \label{eq:beta for ld_1}\\
&\D \ld_2 = 3(\ld_2^2+\ld_{12}^2) \label{eq:beta for ld_2}\\
&\D \ld_{12} = (\ld_1+\ld_2)\ld_{12}+4\ld_{12}^2+8y^2\ld_{12}\label{eq:beta for ld_12}
\end{align}
Before solving the system numerically, let us first obtain some intuition for how the ``off-diagonal" coupling $\ld_{12}$ flows. The whole point of our study is to consider the case for which $\ld_{12}$ is parametrically smaller than the other couplings, so suppose we can drop terms of $O(\ld_{12}^2)$. 
If we drop the $O(\ld_{12}^2)$ term in Eq.~(\ref{eq:beta for ld_12}), then the flow equation for $\ld_{12}$ is of the form $\D\ld_{12} = \F(\mu)\ld_{12}(\mu)$, with $\F = \ld_1+\ld_2+4\ld_{12}+8y^2$, and the equation can be integrated:
\begin{equation}\label{eq:approx sol for ld_12}
\ld_{12}(m) \approx \ld_{12}(\M)\, e^{-\int_m^\M \frac{d\mu}{\mu}\mathcal F(\mu)}\;.
\end{equation}
Thus in the limit $\ld_{12}(\M) \to 0$, the coupling $\ld_{12}(\mu)$ gets renormalized multiplicatively, so perhaps contrary to intuition, we should expect that the assumption $\ld_{12}(\mu) \ll 1$ remains valid even for $\mu \sim m \ll \M$.
\\\\
Now we solve the system of ordinary differential equations \{(\ref{eq:beta function g}), (\ref{eq:beta function y})-(\ref{eq:beta for ld_12})\} numerically. For the high energy scale we take the extreme case $\M \approx 10^{15}$ GeV and we flow to the weak scale, $m \approx 10^2$ GeV, in which case $t \equiv \ln(\mu/\M)$ takes values approximately in the range $0 \geq t \geq -30$.
\\\\
In the SM the strong coupling (in the $\overline{\text{MS}}$ renormalization scheme for concreteness) at the $Z$ pole mass is known to be \cite{pdg qcd review} $g_S(M_Z) \equiv \sqrt{4\pi\al_S(M_Z)} \approx 1.22$ and runs to $\sim 0.5$ at the scale of grand unification in a typical $SU(5)$ model with a desert between the two scales. Thus for illustrative purposes we take the initial condition $g(\M) = 0.5$.
\\\\
The Yukawa coupling in our toy model should be thought of as the case in which the top quark couples to the ``Higgs-like" scalar $\phi_1$. In the model studied in the main text, the $A_4$-triplet Higgs field $\Phi$ is assumed to give mass to both the charged leptons and the quarks, so the situation $y \sim y_t \sim 1$ is the one appropriate for the present study. Thus we take $y(\M) = 1$.
\\\\
For the scalar interactions, the purpose is to study the case $\ld_1(\M) \sim \ld_2(\M) \sim 1$ and\footnote{In a two-Higgs doublet model with Higgs doublets $\phi_1$ and $\phi_2$, the VEVs must align in the charge-conserving vacuum so that the photon does not obtain a mass. This requires the coefficient of the interaction $|\phi_1^\da\phi_2|^2$ to be negative. Furthermore, if $\im(\bra\phi_1\ket\bra\phi_2\ket) = 0$ then the coefficient of $(\phi_1^\da\phi_2)^2+h.c.$ should also be negative.~\cite{2hdm RG} In the present case, the coupling $\ld_{12}$ stands roughly for these two interactions as well as for the coefficient of $|\phi_1|^2|\phi_2|^2$. The running of $\ld_{12}$ will be slow enough such that its sign does not flip, and so if we extrapolate to the $SU(2)\x U(1)$-invariant case we can just pick the appropriate sign at the high energy scale $\M$, which is taken much less than the typical GUT scale, such that electromagnetism remains unbroken at low energy.} $|\ld_{12}(\M)| \ll 1$. To be concrete we take $\ld_1(\M) = \ld_2(\M) = 1$ and $\ld_{12}(\M) = 10^{-3}$.
\\\\
The running couplings $g(\mu), y(\mu), \ld_1(\mu), \ld_2(\mu)$, and $\ld_{12}(\mu)$ are displayed in Figs.~\ref{fig:g},~\ref{fig:y}, \ref{fig:ld1},~\ref{fig:ld2}, and~\ref{fig:ld12}, respectively. The gauge coupling blows up in the deep infrared, as expected, and carries the Yukawa coupling with it into the nonperturbative regime. Both couplings remain $\sim1$ down to $t \sim -25$, or $\mu/\M = e^{\,t} \sim 10^{-11}$. 
\\\\
The quartic self-coupling $\ld_1$ remains essentially fixed to its high-scale value until about the same scale at which $g$ and $y$ become non-perturbative, and the quartic-self coupling $\ld_2$ runs down slowly from $\ld_2(\M) = 1$ to $\ld_2(e^{-25}\M) \sim 0.6$. The difference in running between the two is of course due to the different roles of $\phi_1$ and $\phi_2$ in the model: the gauge coupling feeds into the running for the Yukawa coupling, which contributes only to interactions involving $\phi_1$ and not $\phi_2$. 
\\\\
The main figure of interest is Fig.~\ref{fig:ld12}, which shows that the off-diagonal coupling $\ld_{12}$ runs down from its high-scale value $\ld_{12}(\M) = 10^{-3}$ to ever smaller values, down to $\ld_{12}(e^{-25}\M) \sim 2\times10^{-4}$. Extrapolating these qualitative results to the model in the paper, we expect that the rather mild separation of scales between the weak scale and the ``heavy" mass scales of either the Higgs doublet $\phi$ [Eqs.~(\ref{eq:higgs doublets}),~(\ref{eq:PH mass relation}),~(\ref{eq:PH mass})] or the Higgs triplet $\Del$ [Eqs.~(\ref{eq:higgs triplets}),~(\ref{eq:c_1})] does not spoil the small initial values assumed for the quartic self-couplings in the scalar potential, which would mix the ``charged lepton" and ``neutrino" sectors of the model and introduce the sequestering problem. 
\end{document}